\documentclass[]{sigplanconf}

\usepackage[utf8x]{inputenc}
\usepackage{amsmath}
\usepackage{amsfonts}
\usepackage{verbatim}
\usepackage{moreverb}
\usepackage{listings}
\usepackage{graphicx}
\usepackage{url}

\DeclareUnicodeCharacter{0181}{$\micro$}		
\DeclareUnicodeCharacter{0931}{$\Sigma$}		
\DeclareUnicodeCharacter{8721}{$\Pi$}			
\DeclareUnicodeCharacter{8593}{$\uparrow$}		
\DeclareUnicodeCharacter{8595}{$\downarrow$}	
\DeclareUnicodeCharacter{0915}{$\Gamma$}		
\DeclareUnicodeCharacter{8657}{$\Uparrow$}		
\DeclareUnicodeCharacter{8659}{$\Downarrow$}	
\DeclareUnicodeCharacter{0960}{$\pi$}			
\DeclareUnicodeCharacter{0966}{$\varphi$}		
\DeclareUnicodeCharacter{8901}{$\cdot$}			
\DeclareUnicodeCharacter{8709}{$\emptyset$} 	
\DeclareUnicodeCharacter{0172}{$\neg$}			
\DeclareUnicodeCharacter{8658}{$\Rightarrow$}	
\DeclareUnicodeCharacter{8712}{$\in$} 			
\DeclareUnicodeCharacter{8715}{$\ni$} 			
\DeclareUnicodeCharacter{0956}{$\mu$}			
\DeclareUnicodeCharacter{0963}{$\sigma$}		
\DeclareUnicodeCharacter{0961}{$\rho$}			
\DeclareUnicodeCharacter{0947}{$\gamma$}        
\DeclareUnicodeCharacter{7838}{$\Beta$}         
\DeclareUnicodeCharacter{8747}{$\int$}			

\begin{document}

\conferenceinfo{Technical Report \#UGA-CS-LSDIS-TR-11-011}{Department of Computer Science, University of Georgia, Athens, Georgia (May 2011)}
\copyrightyear{2011}
\copyrightdata{[to be supplied]}

\titlebanner{Draft --- Author's/Reviewer's Eyes Only --- Do Not Distribute}        
\preprintfooter{ScalaTion Unicode Paper for GPCE'11}   

\title{Unicode in Domain-Specific Programming Languages for Modeling \& Simulation}
\subtitle{ScalaTion as a Case Study}

\authorinfo{
	Michael E. Cotterell
	\and John A. Miller
	\and Tom Horton
}{
	Department of Computer Science \\
	University of Georgia \\
	Athens, GA  30602
}{
	mepcott@uga.edu
    \and jam@cs.uga.edu
    \and thorton@uga.edu
}

\maketitle

\begin{abstract}
As recent programming languages provide improved conciseness and flexibility of syntax, the development of embedded or internal Domain-Specific Languages has increased. The field of Modeling and Simulation has had a long history of innovation in programming languages (e.g. Simula-67, GPSS). Much effort has gone into the development of Simulation Programming Languages.

The ScalaTion project is working to develop an embedded or internal Domain-Specific Language for Modeling and Simulation which could streamline language innovation in this domain. One of its goals is to make the code concise, readable, and in a form familiar to experts in the domain. In some cases the code looks very similar to textbook formulas. To enhance readability by domain experts, a version of ScalaTion is provided that heavily utilizes Unicode.

This paper discusses the development of the ScalaTion DSL and the underlying features of Scala that make this possible. It then provides an overview of ScalaTion highlighting some uses of Unicode. Statistical analysis capabilities needed for Modeling and Simulation are presented in some detail. The notation developed is clear and concise which should lead to improved usability and extendibility.

\end{abstract}




\section{Introduction}

ScalaTion is an embedded Domain-Specific Language (DSL) for modeling and simulation (M\&S). M\&S has had a long history of using both General-purpose Programming Languages (GPLs) and Simulation Programming Languages (SPLs). Traditionally, the SPLs may be viewed as external DSLs, although M\&S is broader than many domains of study. Thus, SPLs require many of features of a GPL and the fact that they are external DSLs means that they require extensive custom language support and longer development cycles.

Deursen defines DSLs as ``programming languages or executable specification languages that offer, through appropriate notations and abstractions, expressive power focused on, and usually restricted to, a particular problem domain.''  \cite{deursen:dsl} Providing Unicode support in DSLs is a natural way to facilitate this expressive power by enabling domain-specific notations in programming. DSLs are often implemented using a GPL. The differences between DSLs and GPLs is covered quite extensively by Deursen. \cite{deursen:dsl} Many of these DSLs are implemented externally through the use of lexical parser combinators. However, as covered by Hofer \cite{hofer:dsl, hofer:polymorphic}, there are domain-specific embedded languages (DSELs) in which the DSL is ``embedded as a library into a typed host language instead of creating an external DSL''. These are referred to simply as internal or embedded DSLs.

Until recently, it was difficult to find a GPL suitable for building an embedded DSL for M\&S. Desirable features and capabilities include the following:

\begin{itemize}

\item
\textit{Object-Oriented, Functional Programming Language.}
Mature object models in modern programming languages allow programmers to work with various levels of hierarchy and abstraction. This leads to efficiency and code reuse. Functional programming paradigms help increase readability by emphasizing immutable states and the application of functions as opposed to imperative procedures. It is difficult to find languages that take full advantage of both object-oriented and functional paradigms.

\item
\textit{Support for Unicode.}
Unicode character encoding enables a great number of additional characters outside the traditional ASCII subset to be included in programming languages. Languages can support these characters in literals, identifiers and operators. Although most modern programming languages support Unicode characters in literals, fewer support such characters in their identifiers and operators which greatly diminishes their usefulness in the domain-engineering of an embedded or internal DSL.

\item
\textit{Adequate Performance.}
Some GPLs and their corresponding embedded or internal DSLs suffer a hit when it comes to execution speed due to many reasons. These reasons include the fact that some of them are interpreted languages or that some of them are dynamically, instead of statically, typed. For M\&S, because of its compute intensive nature, an ideal GPL should be both statically typed and allow compilation to machine code (at least using a Just-In-Time (JIT) compiler) in order to minimize overhead and improve the speed of execution of its programs.

\end{itemize}

The Scala Programming Language provides these features and capabilities in a form familiar to Java programmers, so that such programmers can quickly program in Scala. (This has been the case when Scala has been used in classes at the University of Georgia.) An overview of ScalaTion, based on Scala, is given in Miller et al. \cite{miller:scalation}, which highlights the five modelling paradigm (or world-views) supported by ScalaTion (event, activity, process, state, dynamics based modelling).

Just as we chose Scala as our GPL because, among other reasons, it is statically typed and supports Unicode, there appears to be a growing trend toward using Scala for creating DSLs.

\begin{table}[htb]
\begin{center}
\begin{tabular}{|l|l|} \hline
{\bf DSL} & {\bf Domain} \\ \hline \hline
Apache Camel Scala DSL \cite{apache:camel:dsl} & Routing \\ \hline
OptiML \cite{chafi:ml} & Machine Learning \\ \hline
Regions \cite{hofer:polymorphic} & Image Processing \\ \hline
Sake \cite{wampler:sake} & Build Tool \\ \hline
ScalaTion \cite{miller:scalation} & Modeling \& Simulation \\ \hline
Squeryl \cite{squeryl:dsl} & Relational Databases \\ \hline
\end{tabular}
\caption{Examples of DSLs implemented in Scala}
\end{center}
\end{table}

In this paper, we focus on the statistical capabilities provided by ScalaTion.
These capabilities have the following uses in M\&S:

\begin{itemize}

\item
\textit{Random Variate Generation.}
M\&S requires the use of a good random number generator
and several random variate generators for common
probability distributions.
ScalaTion provides the following distributions that mixin
the \texttt{Variate} trait.
Bernoulli,
Beta,
Binomial,
Cauchy,
ChiSquare,
Deterministic,
Discrete,
Erlang,
Exponential,
Fisher,
Gamma,
Geometric,
HyperExponential,
HyperGeometric,
LogNormal,
NegativeBinomial,
Normal,
Poisson,
Randi,
Random,
RandVec,
StudentT,
Triangular,
TruncatedNormal,
Uniform and 
Weibull.

\item
\textit{Output Analysis.}
The purpose of output analysis is to obtain reliable
statistics from simulation runs including point and interval
estimates (e.g., means and confidence intervals).
In this paper, we illustrate how concisely and straightforwardly
the Method of Batch Means can be implemented in ScalaTion.
This method makes one long run and divides it into batches,
so that the batch means are sufficiently uncorrelated
and the confidence interval computed from the batch means and
centered around the grand mean is adequately tight.

\item
\textit{Comparative Analysis.}
M\&S is often used to design systems.
As such it is frequently useful to compare alternative designs.
For example, one factor that could affect the performance
(e.g, response time) of a Database Management System (DBMS) is the
type concurrency control protocol used.
A one-way Analysis of Variance (ANOVA) test could be conducted
to determine if the effect of changing concurrency control
protocols is significant.
One might also wish to determine how the size of the database cache and the speed of main memory affect the response time of the DBMS.
Multiple Regression can be used to address this question.

\end{itemize}

The rest of this paper is organized as follows:
In section 2, we discuss the features and capabilities
that are desirable for an embedded DSL for M\&S,
including the advantages of using an Object-Oriented,
Functional Programming Language as a basis,
the benefits gained by the increasing use of Unicode
in programming languages, and
the performance advantages of using a statically typed,
compiled language.
Section 3 provides an overview of ScalaTion and
highlights some of its features, particularly its
conciseness and use of Unicode.
In section 4, the statistical capabilities of the ScalaTion DSL needed for M\&S are illustrated
with examples.
Section 5 addresses some practical issues that arise
when using Unicode.
Finally, conclusions and future work are given
in section 6.

\section{Embedded DSL for M\&S}

In this section, we discuss the itemized list of language features
and capabilities presented in the Introduction in more detail.

\subsection{GPL Language Features}

GPLs that take advantage of both Object-Oriented and Functional programming language features enable the uses of these features in their embedded or internal DSLs. These features are useful because they allow programmers to work with various levels of abstraction while also increasing readability.

\subsubsection{Object-Oriented, Functional Language Features in Scala}

Many of Scala's object-oriented, functional language features make it ideal for implementing a embedded or internal DSL. These include:

\begin{itemize}

\item \textit{Functional Object Model.} Scala's object model provides the benefits of both object-oriented and functional programming paradigms. In Scala, everything is an object \cite{odersky:spec}.

\item \textit{For Comprehensions, Folds, and Ranges.} More functional language features such as these provide powerful abstractions for writing intuitive segments of code. In most cases, the difference between parallel and sequential version of these operations are simply a matter of the implementation of underlying data structures.

\item \textit{Mutability \& Immutability.} The ability to enforce the immutability of an object helps enforce functional data structures and create code with fewer side effects. This also helps create code that is more thread-safe.

\item \textit{Implicit Conversions.} This language feature enables DSL designers to extend the methods and operations available to core language classes, traits and types. We will examine this particular feature more closely in Section 3.

\item \textit{Tuples.} Built-in functionality for handling tuple types not only helps enforce functional programming paradigms but also aids in statically typed pattern matching.

\item \textit{Generic Arrays.} This particular feature, which is not available in Java \cite{odersky:spec}, enables the construction of generic containers such as vectors and matrices with underlying statically typed arrays. This minimizes the need for casting.

\item \textit{Name-based Operator Precedence.} Although this particular feature stems from Scala's functional object model, we take take advantage of it in our implementation of operators defined with Unicode identifiers.

\end{itemize}

\subsubsection{Statically Typed, Compiled Language Features in Scala}

M\&S is compute intensive for several reasons: systems being modeled are often complex consisting of many subparts, systems are typically simulated over time, simulation runs need to be replicated to obtain reliable statistics, and alternative designs and scenarios need to be considered. Many of Scala's statically typed, compiled language features make it ideal for implementing an embedded or internal DSL. These include:
\begin{itemize}

\item \textit{Typed Language.} Scala enforces the type of variables, and if a parameter specifies a specific type then only that type is allowed. There are instances where a different type will be accepted, but those cases are the result of implicit conversions that are explicitly defined by the programmer.

\item \textit{Static Semantics.}
Scala checks and validates semantic rules, or wellformedness, at compile time \cite{odersky:spec}. Examples of static semantics, in general, include rules (usually defined by some context-free grammar \cite{hofer:polymorphic}) such as identifier declaration and uniqueness in matching labels \cite{scott:programming}. 

\item \textit{Compiled Language.} Scala compiles to bytecode which is executed by the Java Virtual Machine (JVM) at runtime. This speeds up execution when compared to interpreted languages because it removes intermediate steps involved in translating the high-level level language code into machine code.

\end{itemize}

\subsection{Unicode in Programming Languages}

The language one uses to tell a computer what one wants done can have a large impact
on the speed and accuracy of doing this, i.e., writing a computer program.
In the 1960's, there was substantial progress in programming languages, notably
Algol-60, Simula-67 and Algol-68. Since that time, in some sense progress has
been slower, although advances in object-oriented languages and functional languages
has been important.

One barrier to making programming languages or domain-specific languages simultaneously
more concise and more readable, is the limited character set.
The use of Unicode in programming languages allows language designers and programmers to utilize a wider
range of characters than what is contained in the traditional ASCII subset. Recently, languages are
providing ever greater support for Unicode as indicated in the table below:

\begin{table}[htb]
\begin{center}
\begin{tabular}{|l||l|l|l|} \hline
{\bf Use of Unicode} & {\bf Java} & {\bf Scala} \\ \hline \hline
Character Literals & yes & yes \\ \hline
String Literals & yes & yes \\ \hline
Method Names & yes & yes \\ \hline
Prefix Operators & no & no \\ \hline
Infix Operators & no & yes \\ \hline
Postfix Operators & no & yes \\ \hline
\end{tabular}
\caption{Unicode in Programming Languages}
\end{center}
\end{table}

\subsubsection{Support for Unicode in Java \& Scala}

According to the Java Language Specification \cite{gosling:spec} and the Scala Language Specification  \cite{odersky:spec}, both Java and Scala are programming languages which compile to the JVM (Java Virtual Machine) and support Unicode. They support character literals, string literals, and identifiers composed of characters within the Unicode Basic Multilingual Plane (BMP) character set. While this only includes characters in the range 0x0000–0xFFFF, the idea, as described in the Unicode Standard \cite{unicode:standard}, is that this set contains the ``majority of common-use characters for all modern scripts of the world''.

It is interesting to note that the Scala Language Specification also states that infix operators can be defined with any arbitrary, but syntactically legal, identifier. The language even reserves the Unicode equivalents of some built-in operators: `⇒' is equivalent to the `=$>$' operator, and `←' is equivalent to the `$<$-' operator used in comprehensions. The existence of these operators illustrate that there already exists an interest in representing operators with their Unicode equivalents. The precedence of these and other infix operators in Scala is determined by the identifier's first character. In Scala, Unicode characters are considered to be ``special characters'' and have a higher precedence than all other operators.

\begin{table}[htb]
\centering
\begin{tabular}{l}
(all letters) \\
$|$  \\
\^{} \\
\& \\
$<$ $>$ \\
$=$ ! \\
: \\
$+$ $-$ \\
* / \%{} \\
(all other special characters) \\
\end{tabular}
\caption{Operator Precedence in Scala (Low to High) \cite{odersky:spec}}
\end{table}

Here is a toy example, written in valid Scala, demonstrating an infix operator defined with a Unicode character. Note that we accomplish this by simply defining a new method `\texttt{≡}' that takes a rational number as a parameter and returns whether or not they are equivalent. This is possible for two reasons; according to Scala's unified object model \cite{odersky:overview}, every operation is the invocation of a method; the equality test using the == in Scala compares objects by value and not by reference. \cite{odersky:overview}

\begin{listing}{1}
case class Rational(a: Int, b: Int) {
  def ≡ (that: Rational)
    = (a/b) == (that.a/that.b)
}

val (x, y) = (Rational(1, 2), Rational(2, 4))
println("x == y = " + (x == y)) // false
println("x ≡ y  = " + (x ≡ y)) // true
\end{listing}

Another interesting thing to note is the way in which Scala handles postfix functions, methods, and operators. Although Scala, like Java, supports a dot-syntax for calling such methods, using a dot or period between the object identifier and the method identifier is not required. When a method only takes a single argument parameter, the language also permits the omission of parentheses around the parameter identifier. In both cases, the dot or the opening parentheses must usually be replaced with at least one space character. In fact, this is what makes infix operators work the way they do. Infix operators are merely defined as a methods which only take a single argument parameter. However, experimentation with Scala 2.8.1.final and 2.9.0.final indicates that a space between two identifiers may also be omitted under special circumstances. Mainly, the characters that meet where the space would usually occur must have different precedences. As Unicode characters are considered ``special characters'' in terms of precedence, this makes it possible to implement some interesting postfix Unicode operators.

As a toy example of how Scala permits the omission of a space between an object identifier and a postfix operator defined with a Unicode character, consider the following implementation of a case class \texttt{Num} for integral numbers. This class simply encapsulates an \texttt{Integer} and provides a Unicode postfix operation that returns the square of the \texttt{Integer} value.

\begin{listing}{1}
case class Num(n: Int) {
  def ²() = Num(n * n)
}

val a = Num(2)
val square = a²
println(square) // output: Num(4)
\end{listing}

\noindent This example is not without one critical limitation as it only allows for single-digit exponents.

Although the application of Unicode prefix operators would be interesting, unfortunately, the Scala Language Specification only permits the following identifiers to be used as prefix operators: `$+$', `$-$', `$!$', and `$\sim$'. Perhaps support for additional prefix operators will be added in future versions of Scala.

\subsubsection{Support for Unicode in DSLs}

As Unicode enables the use of an extended character set (extended in the sense that it provides more characters than ASCII), DSLs that support Unicode have the opportunity to go beyond simply using domain-specific terminology and provide domain-specific notation. Special symbols for functions and operators can be defined in a DSL that are familiar to domain-experts and provide concise, easily-readable code. Such support is automatically available to embedded or internal DSLs that are implemented using Unicode-supported GPLs. An example of a small, toy DSL for Boolean Algebra, implemented by Gabriel C. \cite{gabriel:dsl}, demonstrates that Unicode operators can be used to simplify the notation of an embedded or internal DSL.  

\section{Overview of the ScalaTion DSL}

Our case study on Unicode in domain-specific programming languages will focus on ScalaTion, an embedded or internal DSL implemented in Scala which supports multi-paradigm simulation modeling including dynamics, activity, event, process and state oriented models.

As ScalaTion is a DSL implemented in Scala, it benefits from all the language features mentioned earlier.
Anything that can be done in Scala can be done in ScalaTion.

\subsection{Techniques for Adding Unicode Support}

With respect to external DSLs, Unicode support can be added by adjusting the grammar and using parser combinators. However, when it comes to adding Unicode support to embedded or internal DSLs, the language designer must adhere to the language constructs of the GPL. In the case of ScalaTion, the addition of Unicode support was achieved in three different ways as discussed in the next three subsections.

\subsubsection{Defining new Classes and Objects}

The first and most straightforward way in which we added Unicode support to ScalaTion is through the creation of new classes and objects. Consider the \texttt{Vec} class, an implementation of a numeric vector, which defines and implements several operators including the dot product operator.

\begin{listing}{1}
val vec1 = Vec(1, 2, 3)
val vec2 = Vec(4, 3, 2)

val dp1 = vec1 dot vec2
println(dp1) // output: 16

val dp2 = vec1 ⋅ vec2
println(dp2) // output: 16
\end{listing}

As Scala's functional object model allows us to define operators as methods, the example above is implemented
by merely adding a definition for the dot product method in \texttt{Vec}.

\begin{listing}{1}
class Vec [T] extends ... {
  def ⋅ (rhs: Vec[T]) = ... // implementation
  def dot (rhs: Vec[T]) = this ⋅ rhs
}
\end{listing}

\subsubsection{Mixin Compositions}

Another way we added Unicode support to ScalaTion is through mixin compositions. According to Odersky \cite{odersky:2003:nominal}, mixins provide classes and objects with members which can be referred to from other definitions in the class. In Scala, when mixins are defined independently of type instantiations they are called traits. This is somewhat similar to idea of interfaces and abstract base classes in other languages such as C++ and Java.

ScalaTion provides a trait called ``\texttt{ScalaTion}'' that can be mixed-in with any Scala object or class in order to provide that object or class with some of the basic functionality implemented in the DSL. This is done, for example, when a Scala application wants to use the ScalaTion DSL. Here is an example of how to do this in  Scala 2.9.0.final:

\begin{listing}{1}
object ExampleApp
  extends App with ScalaTion {
  val s = Vec(1, 2, 3)
  val test = 2 ∈ s
  val sum = Σ(s)
  println(test) // output: true
  println(sum)  // output: 6
}
\end{listing}

\noindent As can be seen, mixing-in the \texttt{ScalaTion} trait enables the \texttt{ScalaApp} object to utilize the types and operators provided by the DSL. This makes it easier to utilize language extensions as compared to Java alternatives that involve static inputs.

\subsubsection{Implicit Conversions}

The third way we added Unicode support to ScalaTion is through implicit conversions. According to Odersky \cite{odersky:2006:poor}, one can define a unary function from type S to type T, labelled implicit, that will provide a conversion from one type to another under certain contexts. To make this clearer, let us look at how support for the infix Unicode operator `∈' was added in our previous examples. Inside the \texttt{ScalaTion} trait, we define an implicit conversion from type \texttt{Any} to type \texttt{RichAny}.

\begin{listing}{1}
implicit def any2RichAny(elem: Any)
  = new RichAny(elem)
\end{listing}

\noindent In this example, Scala's scoping rules will enable users of the DSL to call the methods and operations defined in \texttt{RichAny} with \texttt{Any} type. This allows us to extend the functionality of the GPL in the embedded or internal DSL.

\subsection{Basic Code Samples}

Mixing-in the \texttt{ScalaTion} trait with classes and objects provides them with many useful constants and functions that are defined using Unicode characters. In this section, we present some of the functions that benefit syntactically from Unicode.

\subsubsection{Exponentiation}

ScalaTion's DSL provides a Unicode infix operator for exponentiation. The character that we chose is the up-arrow, which is consistent with a general implementation of Knuth's up-arrow notation \cite{knuth:math} and provides the correct precedence level (although it does not have right to left associativity).

\begin{listing}{1}
val exp1 = 2↑2     // 4
val exp2 = 2↑2↑2   // 16
\end{listing}

\noindent We decided to be consistent in our representation of roots. The \texttt{ScalaTion} trait also provides a Unicode infix operator for calculating $n^{th}$-roots of a number using down-arrow notation.

\begin{listing}{1}
val root1 = 4↓2           // 2
val root2 = 4↓2↓2         // 1.41421...
val test  = 4↑0.5 == 4↓2  // true
\end{listing}

\subsubsection{Factorials}

We chose to support factorials in a traditional way, using the exclamation mark as a postfix operator. This operator may be used on any \texttt{Numeric} type.

\begin{listing}{1}
val fac1 = 4!       // 24
val fac2 = 3.5!     // 11.63172...
\end{listing}

\noindent The \texttt{ScalaTion} trait also provides Unicode infix operators for the rising and falling factorials. We chose the double up and down arrows for this notation.

\begin{listing}{1}
val rising = 4 ⇑ 4  // 840
val falling = 4 ⇓ 4 // 24
\end{listing}

\subsubsection{Product Series}

Support for product series is also included in the \texttt{ScalaTion} trait. We implemented this functionality using a Unicode method identified by the unary product symbol which is similar to the Greek capital letter `∏'. The range is defined using Scala's built-in \texttt{Range} class.

\begin{listing}{1}
val prod1 = ∏(1 to 3)                // 6
val prod2 = ∏(1 to 3, i ⇒ i↑2)      // 36
\end{listing}

\noindent Product series are also defined for sets, vectors, and other indexed sequences.

\begin{listing}{1}
val set = Set(1, 2, 3, 4)
val prod3 = ∏(set)                   // 24
val prod4 = ∏(0 to 2, i ⇒ set(i))   // 6
\end{listing}

\subsubsection{Summation Series}

Support for summation series is also included in the \texttt{ScalaTion} trait. We implemented this functionality using a Unicode method identified by the unary sum symbol which is similar to the traditional Greek capital letter `Σ'. The range of the summation is defined using Scala's built-in \texttt{Range} class.

\begin{listing}{1}
val sum1 = Σ(1 to 3)                 // 6
val sum2 = Σ(1 to 3, i ⇒ i↑2)       // 14

val set = Set(1, 2, 3, 4)
val sum3 = Σ(set)                    // 10
val sum4 = Σ(0 to 2, i ⇒ set(i))    // 6
\end{listing}

\subsubsection{Definite Integrals}

Support for definite integrals is also included in the \texttt{ScalaTion} trait. We implemented this functionality using a Unicode method identified by the integral symbol. The range of the integral is defined using Scala's built-in \texttt{Range} class.

\begin{listing}{1}
val int1 = ∫(1 to 4, i ⇒ i↑2)        // 21
\end{listing}

\subsubsection{Sets and Set-like Objects}

Support for various set operations is also included in the \texttt{ScalaTion} trait. Many of these operations work with other \texttt{SetLike} objects as well. Some even work with sequences. For the examples below, let us consider the following two sets:

\begin{listing}{1}
val set1 = Set(1, 2, 3, 4)
val set2 = Set(1.0, 2.0, 3.0, 4.0)
\end{listing}

\noindent Testing membership in a set is as easy the following.

\begin{listing}{1}
println(2 ∈ set1)   // output: true
println(5 ∈ set1)   // output: false
\end{listing}

\noindent Also, as these operations are statically typed, conversions from one type to another are implicitly performed on a contextual basis. For example, in Scala, a \texttt{Int} can be implicitly converted to a \texttt{Double}. This enables the following statement.

\begin{listing}{1}
println(2 ∈ set2)   // output: true
\end{listing}

\noindent The benefits of Scala's type system prevent certain kinds of errors. For example, the following code will not compile due to a static type error.

\begin{listing}{1}
println(2 ∈ set3)   // will not compile
\end{listing}

ScalaTion also supports the universal and existential quantifiers.

\begin{listing}{1}
println(∀(set1, _ > 2))  // output: false
println(∃​(set1, _ > 2))  // output: true
\end{listing}

\subsubsection{Numeric Vectors}

Support for numeric vectors and their operations are included in the \texttt{ScalaTion} trait. As mentioned earlier, in addition to other common vector operations, the \texttt{Vec} class defines a Unicode infix operator for the dot product.

\begin{listing}{1}
val vec1 = Vec(1, 2, 3)
val vec2 = Vec(4, 3, 2)

val dp1 = vec1 ⋅ vec2
println(dp1) // output: 16
\end{listing}

As with \texttt{Array} or \texttt{IndexedSeq} in Scala, the \texttt{Vec} class provides random access to the vector elements using the \texttt{apply} method. ScalaTion overloads this method to accept a \texttt{Range} parameter as well. When this is done, a new vector is returned that is equivalent to the original vector sliced at the bounds of provided range.

\begin{listing}{1}
val v = Vec(1, 2, 3, 4, 5, 6, 7)

val v2 = v(2)           // 3
val v3 = v(3)           // 4

val v2_4 = v(2 to 4)    // Vec(3, 4, 5)
val v2_3 = v(2 until 4) // Vec(3, 4)
\end{listing}

\noindent In an attempt to provide a more mathematically-recognized notation for Scala's built-in \texttt{Range}, we implemented an alias for the \texttt{to} method in the form of a short ellipses (using the Unicode ellipse character). This enables users of ScalaTion to write the following:

\begin{listing}{1}
val v2_4 = v(2...4)     // Vec(3, 4, 5)
\end{listing}

\subsection{Unresolved Issues}

Choosing Scala as ScalaTion's GPL was not without some limitations. For example given our current approach, we are unable to implement the following language features:

\begin{itemize}
\item \textit{Unicode prefix operators.}
As mentioned earlier, Scala does not allow custom prefix operators in ASCII, Unicode, or otherwise. This required us to implement prefix operators as ordinary methods and functions, requiring parentheses.

\begin{listing}{1}
val b = true
println(¬(b)) // output: false
\end{listing}

\item \textit{Control of precedence levels.} As groups of operators in Scala have a fixed level of precedence and all Unicode-identified operators fall into the same group, we are unable to differentiate the precedence of such operators when evaluated. In order to get around this problem, we would could take two approaches. First, we could perform some sort of text-substitution preprocessing that replaces Unicode operators with operators of appropriate precedence. For example, in order to preserve the precedence of set union and intersection, we could replace the $\cup$ and $\cap$ characters with \texttt{$|$} and \texttt{\&} respectively. In order to provide the subset operator with a lower precedence, we could replace the $\subseteq$ character with \texttt{subsetOf}. Second, we could extend the parser combinators in Scala's compiler through plugins in order to assign precedence without the need for preprocessing. However, both of these approaches are outside the scope of internal or embedded DSLs, because they externalize the language by either increasing the number of intermediate steps required by the end user during the compilation process and by requiring more than just the GPL and DSL library. (There has been some discussions about either including a certain subset of Unicode operators into Scala with different precedent levels or providing full user control over operator precedence \cite{nabble}.)

\end{itemize}

\section{Statistical Analysis using the ScalaTion DSL}
Results are produced in simulation by making multiple runs of a program or by dividing a long run into multiple batches. Each of these produces sample data points that must be analyzed statistically. Consequently, simulation relies heavily upon statistics. The goals guiding the development of the statistical analysis capabilities of ScalaTion are the following:

\begin{itemize}

\item
Make the code concise and intuitive so that someone reading
a Modeling and Simulation or Statistics textbook would find
the code easy to use (not exactly the formulas in the textbook,
but similar enough to be easily recognized).

\item
Make the code reasonably efficient.
Following notation in a textbook too closely may lead to
inefficient code, but if the efficiency leads to
obfuscation, efficiency needs to take a back-seat.

\item
Rely heavily on the use of vectors as this leads to
concise and readable code, and provides opportunities
for parallel processing based on Scala 2.9's
parallel collection classes.

\end{itemize}

\subsection{Random Variate Generation}

ScalaTion provides classes for producing many different kinds of probability distributions. 

\begin{listing}{1}
val rv = Normal(μ, σ)
val x  = rv.gen
\end{listing}

\noindent The \texttt{RandVec} class provides a way to generate a numeric vector populated with a Random Distribution of numbers. Each number in the vector has an associated probability.
By default, all numbers in a \texttt{RandVec} have equal probability. This class extends \texttt{Vec} and therefore supports all of the vector operations discussed earlier. It also mixes-in the \texttt{Variate} trait in order to allow interaction with certain statistical functions.

\subsection{Output Analysis}
Output analysis is the examination of data generated through a simulation. According to Banks, Carson, Nelson and Nicol \cite{bcnn:2000}, the purpose of of the statistical analysis is to estimate the confidence interval or to the number of observations required to achieve a confidence interval. ScalaTion includes a collection of statistical procedures for analyzing the observed variance in a particular variable or series of variables.

Using the mathematical notation described earlier, ScalaTion is able to provide many statistical formulas that look similar to way they are defined in a textbook. In this section, we will present some of the statistical functions provided by ScalaTion that benefit from the use of Unicode in their function identifiers. 

\subsubsection*{Mean and Expectation}

In ScalaTion, the mean of a vector is provided by the \texttt{mean} function. This makes is easier for us to define the mean statistic, μ, for any given \texttt{RandVec}. In the case of the mean, both the sample statistic and the population characteristic are the same.

\begin{listing}{1}
def μ (x: RandVec) = x.mean
\end{listing}

\noindent We should also note that, in general, as the expected value E[x] is also referred to as the mean, μ, or the first moment of x. 

\begin{equation}
\nonumber \mu(x) = E[x]
\end{equation}

\noindent { }

\noindent This, however, is just a matter of abstraction.
In ScalaTion, all types that mixin \texttt{Variate} must define their own mean.

\subsubsection{Mean Square}

The mean square, or second moment, of x is simply the average of the squares of x.

\begin{equation}
\nonumber ms = \mu_2 = \mu(x^2)
\end{equation}

\noindent { }

\noindent In ScalaTion, we define the \texttt{ms} function to take a \texttt{RandVec} and return this value.

\begin{listing}{1}
def ms (x: RandVec) = μ(x↑2)
\end{listing}

\subsubsection{Variance}

In statistics, variance measures how far a set of numbers are spread out from each other. In Banks, Carson, Nelson and Nicol \cite{bcnn:2000}, an equation for population variance is provided. For our purposes, we define this equation using the mean of a \texttt{RandVec} as its expected value. We also utilize the mean square calculation.

\begin{align*}
\nonumber \sigma2 = V(x) &= \mu\left((x - \mu(x))^2\right) \\ 
&= \mu(x^2) - \mu(x)^2 \\
&= ms - \mu(x)^2
\end{align*}

\noindent { }

\noindent This produces the following definition in ScalaTion.

\begin{listing}{1}
def σ2 (x: RandVec) = ms(x) - μ(x)↑2
\end{listing}

\noindent We also define the sample variance \texttt{$\sigma$2\^}.

\begin{listing}{1}
def σ2^ (x: RandVec) = {
  val n = x.dim
  n * σ2(x) / (n-1)
}
\end{listing}

\noindent Note, several sample statistics are provided in ScalaTion, but in this paper we focus on population characteristics for simplicity.

\subsubsection{Standard Deviation}

Standard deviation shows how much variation there is from the mean or expected value.
In Banks, Carson, Nelson and Nicol \cite{bcnn:2000}, it is defined as:

\begin{equation}
\nonumber \sigma = \sqrt{V(x)}
\end{equation}

\noindent { }

\noindent In ScalaTion, we define the standard deviation of a \texttt{RandVec} using the following function definition in Scala.

\begin{listing}{1}
def σ (x: RanVec) = σ2(x)↓2
\end{listing}

\subsubsection{Skewness}

The skewness statistic is a measure of the asymmetry of the probability distribution of a \texttt{Variate}, defined as follows:

\begin{align*}
\nonumber
\mu &= \mu(x) \\
\mu_3 &= \mu(x^3) \\
\sigma &= \sigma(x) \\
\gamma1 &= \frac{\mu_3 - 3 \mu \sigma^2 - \mu^3}{\sigma^3} \\
\end{align*}

\noindent Here is the corresponding definition in ScalaTion:

\begin{listing}{1}
def γ1 (x: RandVec) = {
  val (μ, μ3, σ) = (μ(x), μ(x↑3), σ(x))
  (μ3 - 3 * μ * σ↑2 - μ↑3) / (σ↑3)
}
\end{listing}

\subsubsection{Covariance}

The covariance statistic is a measure of how much two variables change together. It is defined by the following equation.

\begin{align*}
\nonumber
cov(x, y) &= \mu\left((x - \mu(x))(y - \mu(y))\right) \\
&= \mu(xy) - \mu(x) \mu(y)
\end{align*}

\noindent { }

\noindent Here are the corresponding definitions in ScalaTion:

\begin{listing}{1}
def cov (x: RandVec, y: RandVec)
  = μ(x*y) - μ(x) * μ(y)
\end{listing}

\subsubsection{Correlation}

The population correlation statistic, also known as the Pearson product-moment correlation coefficient, is a measure of dependence between two \texttt{Variate} objects. It is defined by the following equation.

\begin{align*}
\nonumber
\rho(x, y) &= \frac{cov(x, y)}{\sigma(x) \sigma(y)}
\end{align*}

\noindent { }

\noindent Here is the corresponding definition in ScalaTion:

\begin{listing}{1}
def ρ (x: RandVec, y: RandVec)
  = cov(x, y) / (σ(x) * σ(y))
\end{listing}

\subsubsection{Autocorrelation}

The first-order autocorrelation statistic of a \texttt{Variate} is the correlation statistic between different ranges of the \texttt{Variate}. We generalize this into the following formula.

\begin{align*}
\nonumber
\rho(x_{0...n-1}) &= \rho(x_{0...n-2}, x_{1...n-1})
\end{align*}

\noindent { }

\noindent In ScalaTion, autocorrelation is defined similarly using the correlation function we defined earlier. When we combine this with ScalaTion's built-in vector slicing, we are able to produce the following Scala code.

\begin{listing}{1}
def ρ (x: RandVec) = {
  val n = x.dim
  ρ(x(0...(n-2)), x(1...(n-1)))
}
\end{listing}

\subsubsection{Batch Means}

ScalaTion extends \texttt{RandVec} to create \texttt{BatchVec} a class for calculating the batch means and confidence levels for simulation. The method of batch means is a popular output analysis technique used for steady-state simulations \cite{bcnn:2000}.

Given an initial batch size of \texttt{b}, we try new batch sizes (doubling for each attempt) until the autocorrelation of the batch means drops to a threshold of, for example, 0.1. In ScalaTion, we implement this using the following Scala code:

\begin{listing}{1}
def makeBatch (b: Int, n: Int = 1): RandVec 
  = // simulate to collect b*n sample data points

def μBatch(x: RandVec, b: Int): RandVec = {
  val n = x.dim / b
  for (i ← 0 until n-1) 
    yield μ(x((i*b)...(i+1)*b-1))
}

def formBatches (b: Int = 10, n: Int = 10,
  x: RandVec = RandVec.ofLength(0)): 
  (Int, RandVec, RandVec) {  
  val y = x ++ makeBatch(b, n)  
  val μVec = μBatch(y, b)
  (ρ(μVec) > 0.1) match {
    case true  => formBatches(2*b, n, y)
    case false => (b, y, μVec)
  }
}
\end{listing}

Now that the batch means are sufficiently uncorrelated, we can compute a confidence interval and determine relative precision (ratio of the confidence interval half-width to the grand mean).

\begin{listing}{1}
var (b, x, μVec) = formBatches()
var (gμ, precision) = (0.0, 0.0)
do {
  (gμ, precision) 
    = (μ(μVec), μVec.interval() / gμ)  
  if (precision > 0.2) {
    x = x ++ makeBatch(b)
    μVec = μBatch(x, b)  
  }
} while (precision > 0.2)
\end{listing}

\noindent The loop above will cause additional batches to be collected until a sufficient relative precision is obtained.

\subsection{Comparative Analysis}

In simulation, comparative analysis may be used to consider design alternatives, e.g., which server configuration is more efficient, two slower, less costly chips or one faster, more expensive chip. Again, as simulation results are stochastic, it is important to use rigorous statistical techniques to compare alternatives. There are several techniques for comparing design alternatives including paired-t tests and ANOVA as well as advanced techniques for ranking and selection \cite{lk:1982}.

\subsubsection{One-way Analysis of Variance}
ScalaTion provides an \texttt{Anova} class and object for performing a one-way Analysis of Variance (ANOVA). One-way ANOVA is often used to compare multiple treatments (e.g., design alternatives) typically using a \texttt{Fisher} distribution. An \texttt{Anova} object can be constructed with either a numerical matrix or a sequence of numerical vectors. For the following examples, let \texttt{m} and \texttt{n} be the dimensions of the input matrix \texttt{x}.

\begin{listing}{1}
val m = x.dim1 // m rows
val n = x.dim2 // n columns
\end{listing}

\noindent Each row of the matrix corresponds to a treatment and contains \texttt{n} replicates.

\subsubsection*{Grand Mean}

The grand mean is the mean of the means of each group \cite{everitt:dict}. It is defined by the following equation.

\begin{equation}
\nonumber
g\mu = \frac{\sum_{i=0}^{m-1} \mu(x_i)}{m}
\end{equation}

\noindent { }

\noindent In ScalaTion, we can define the grand mean using the same formula.

\begin{listing}{1}
def gμ = Σ(0, m-1, i => μ(x(i))) / m
\end{listing}

\subsubsection*{Total Sum of Squares}

The total sum of squares can be written as the sum of the squares of the group deviations. It is defined by the following equation.

\begin{equation}
\nonumber
ss_t = \sum\limits_{i=0}^{m-1} \sum_{j=0}^{n-1} (x_{i,j}^2) - m \cdot n \cdot g\mu^2
\end{equation}

\noindent { }

\noindent In the \texttt{Anova} class, we define the total sum of squares using the same formula.

\begin{listing}{1}
def sst = Σ(0, m-1, i ⇒ Σ(x(i)↑2)) - m*n*gμ↑2
\end{listing}

\noindent The code above takes advantage of applying the exponentiation operator to each element of the vector.

\subsubsection*{Between-groups Sum of Squares}

The between-groups sum of squares can be written as the square of the sum of deviations between each group. It can be defined by the following equation.

\begin{equation}
\nonumber
ss_b = \sum\limits_{i=0}^{m-1}\sum_{j=0}^{n-1} \left(\frac{x_{i,j}^2}{n}\right) - m \cdot n \cdot g\mu^2
\end{equation}

\noindent { }

\noindent We implement this formula in ScalaTion using the following Scala code.

\begin{listing}{1}
def ssb = Σ(0, m-1, i ⇒ Σ(x(i))↑2/n) - m*n*gμ↑2
\end{listing}

\subsubsection*{Within-groups Sum of Squares}

The within-groups sum of squares can be written as the square of the sum of deviations within each group. It can simplified into the difference between the total and between-groups sum of squares. It is defined by the following equation.

\begin{equation}
\nonumber
ss_w = ss_t - ss_b
\end{equation}

\noindent { }

\noindent We easily define this statistic in ScalaTion using the following Scala code.

\begin{listing}{1}
val ssw = sst - ssb
\end{listing}

\subsubsection*{F-statistic}

The F-statistic is the ratio of the between-groups and within-groups sum of of squares divided by their respective degrees of freedom. It is used in conjunction with a Fisher distribution to determine if the values are statistically significant for some probability.

\begin{equation}
\nonumber
f = \frac{ss_b / m-1}{ss_w / m \cdot (n-1)}
\end{equation}

\noindent{ }

\noindent In ScalaTion, we define this value in the \texttt{Anova} class using similar notation.

\begin{listing}{1}
def f = (ssb / m-1) / (ssw / m*(n-1))
\end{listing}

\section{Practical Issues in Using Unicode}

Unicode support for embedded or internal domain-specific languages must include proper tooling.  By this, we mean that adequate tools should be provided so that end users of the language can utilize the special Unicode features. As mentioned in the previous section, there are a few missing capabilities that can make this task difficult. However, they is nothing that cannot be dealt with by extending existing technologies.

\subsection{Input Methods for Unicode}

Any claim of advantages or benefits to our efforts to enhance an internal or embedded domain-specific languages through the use of the extended character set available through Unicode could be legitimately criticized if the user is not able to easily incorporate these characters into their programming environment. This issue is not new and validly applies whenever the difficulties or disadvantages of inclusion of a new feature outweigh the potential benefits. It is for this reason that we elaborate on this problem and some of the methods and technologies that eliminate or greatly reduce end user burden when entering Unicode characters. As with the definition of Unicode itself, this is a dynamic process and the advancement of this goal is ongoing. 

The Unicode Standard - Version 6.0 - Core Specification is a 670 page document that contains
the ``universal character encoding, extensive descriptions and a vast amount of data how the characters function'' \cite{unicode:standard}. The use of Unicode characters and symbols addresses the challenge of computer system users worldwide to expand upon the base problem of being able to utilize characters and symbols beyond that found on a traditional 80 to 100 key keyboard. The breadth of the problem that Unicode addresses can be illustrated by knowing the that Unicode character set covers over 100,000 characters in 93 scripts \cite{unicode:standard}, although we focus on the BMP Unicode subset.

In order for users to effectively utilize the advantages interoperability between different application implementations and the world’s languages that Unicode addresses there must be effective and easy to use methods for entering Unicode characters into a computer system.

The data entry methods currently used seem to fit into the categories of supported by hardware, software, or a combination. Keyboards of many designs have long been used to implement spoken and computer languages. The number of national language keyboards  exceeds 100 different keyboards. There are obvious and known problems with keyboards having a relatively small number of physical keys addressing certain spoken languages.

The BIOS of some systems have been designed to accommodate the limited number of special
keyboard characters. The “Control-Alt-Delete” key sequence (or chord) has been part of
computer users’ entry repertoire for decades .

Similarly, some computer programming languages and applications use symbols that are not
universally found on computer keyboards. The concurrent use of multiple keys, or a chord, is
used to address entry of characters not otherwise found on a keyboard. Certainly, the concurrent
use of a shift key allows for entry of upper and lower case letters. Likewise, the “alt”, “ctrl”,
and “alt-ctrl” keys expand the base keyboard character sets.

An example of one form of special and unique hardware support of an expanded character set is
illustrated by the Art Lebedev Studio keyboard offerings. Their “Optimus Tactus keyboard
does not have physical individual keys removing restrictions upon the shape and size” of keys \cite{lebedev:maximus}.

\begin{figure}[h]
\centering
\includegraphics[scale=0.20]{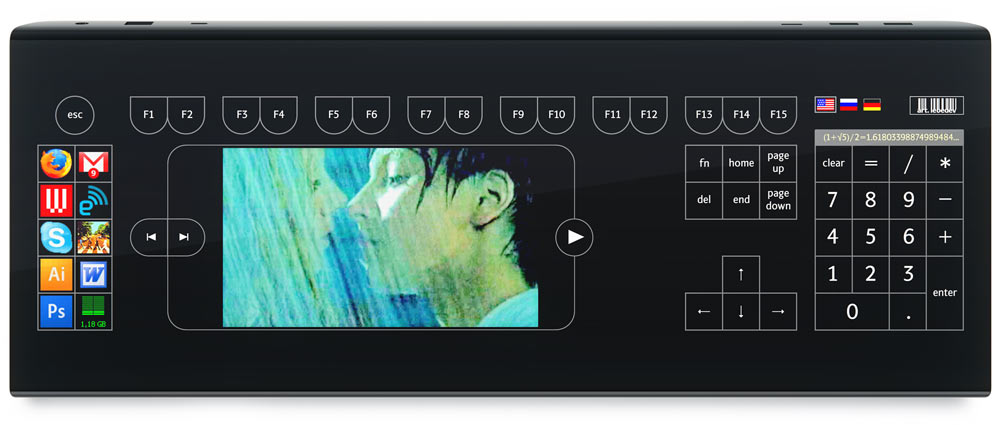}
\caption{Optimus Tactus Keyboard by Art Lebedev Studio}
\end{figure}

\noindent Additionally, any part of the keyboard surface can be programmed to perform a function or to display an image. The “Tactus” can be programmed to appear as a typical qwerty
keyboard or a video image \cite{lebedev:tactus}. The “Maximus” keyboard does have typical physical
keys, but is able to be programmed to enter characters of many languages, special symbols,
HTML code, and math functions. Each key top is a small display indicating what the button is
programmed to do \cite{lebedev:maximus}.

\begin{figure}[h]
\centering
\includegraphics[scale=0.25]{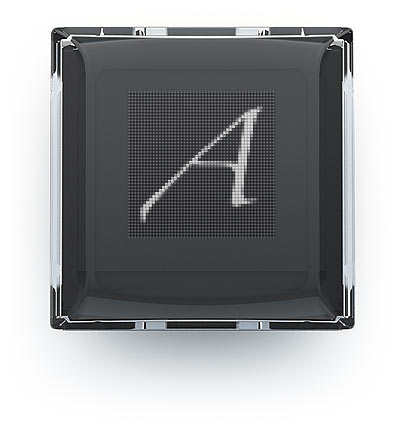} \includegraphics[scale=0.25]{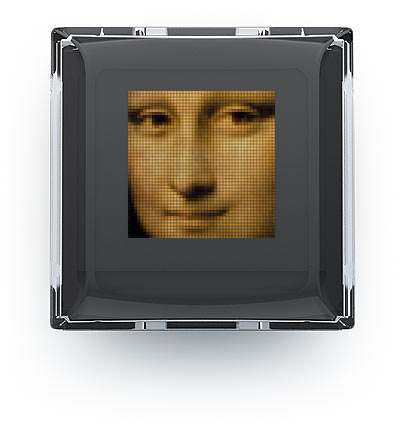}
\caption{Optimus Maximus Keys by Art Lebedev Studio}
\end{figure}

Another example of hardware supporting an extended character set is provided by the X-Keys product series \cite{pi:xkeys}. The X-Keys product series (keyboards, keypads, and other devices) are physical extensions or auxiliary data entry devices. Without a major interruption to the data entry proficiency, a user is able to switch from the standard keyboard to an auxiliary key device thereby utilizing a greatly expanded character set.

\begin{figure}[h]
\centering
\includegraphics[scale=0.20]{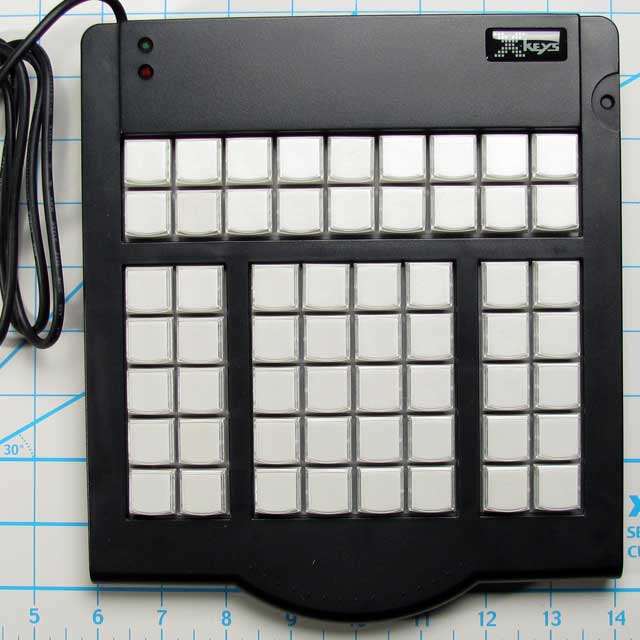}
\caption{XK-Professional by P.I. Engineering}
\end{figure}

Specialty hardware is certainly not a requirement for the entry of Unicode characters. There
are a large (and growing) number of software products that address the data entry of expanded
character sets. Microsoft Windows provides a basic method for entry of Unicode characters.
Through the use of the “alt+num pad” users can enter the Unicode-generated (UTF-16)
character \cite{msdn:altnumpad}. This expands the data entry character set via a standard keyboard. Other software
tools incorporate the combined use of a keyboard and the screen or display. ISO 14755 refers to
this as a screen-selection entry method \cite{iso:14755}.

\subsection{ScalaTion-specific Considerations}

The need for entry of an expanded character set in ScalaTion is not as broad as the generic Unicode character entry problem. Currently, ScalaTion shows how by using a subset of the familiar mathematical symbols produces code that appears closer to the end-user’s problem formulation. It is very important to enable the user an easy and proficient way to enter the Unicode symbols implemented in this version of ScalaTion.

End users have their choice of data entry methods: hardware, software, or combination method. What is important is for the users to understand that data entry should not be considered a stumbling block toward the use of a problem solving tool that uses a character set beyond that of the standard keyboard.

We believe that as hardware and software continues to mature that data entry of expanded
character set will likely include graphics, multi-media, mobile devices, and the full range of input and output devices.

\section{Conclusions and Future Work}

We have developed and presented ScalaTion, an embedded or internal DSL for M\&S, which we believe will streamline language innovation in this domain through its utilization of Unicode-encoded identifiers. The code and documentation is available on the ScalaTion project website: \url{http://code.google.com/p/scalation/}.

Through our case study on ScalaTion, we have demonstrated that there are ways to make Scala code more concise, readable, and in a form more familiar to (M\&S) domain-experts. ScalaTion provides Unicode-identified functions and operators that are easily recognized by domain-engineers in M\&S. Such domain-specific notation enables concise, easily readable code to be written by such engineers and other users of ScalaTion.

We took advantage of three different methods for adding Unicode support to the ScalaTion DSL. The first and easiest method was the creation of new classes and objects that define their own Unicode operators (e.g., the dot product operator in \texttt{Vec}). The second method was through Scala's mixin compositions which enabled us to add Unicode  constants, functions, and operators to the scope of any newly created object. For example, when extending the \texttt{App} trait for easy application creation, we can mixin the \texttt{ScalaTion} trait, which enables the use of these Unicode definitions within the application object. The third method by which we added Unicode support to ScalaTion was through implicit conversions. This enabled us to implicitly add Unicode functions and operators to existing Scala classes, objects, and types (e.g., adding the $\in$ operator to all types, enabling us to test whether they are contained within a set). These methods demonstrate how easy it is to add Unicode support to both new and existing DSLs implemented in the Scala Programming Language.

Although ScalaTion already provides many of the functions needed for programming in M\&S, there is always room for improvement. Here are some of our proposals for future work.

\begin{itemize}
\item \textit{IDE Plugins and Frontends.} We will work on integrating tools for using the ScalaTion DSL into Integrated Development Environments (IDEs). This will include such things as toolbars for selecting Unicode identified operators and extensions to content-assist services for looking up and suggesting operators that are available and contextually relevant. Popular IDEs that currently support Scala via plugins include Eclipse and IntelliJ. Some work has already begun on extending Eclipse to support the ScalaTion DSL via toolbar plugins. In the future, such developments may lead to a unified frontend for ScalaTion similar to the frontends of external DSLs like R, Maple, Mathematica, and MATLAB. Such work will also help ease the input and output of mathematical notations in ScalaTion.

\item \LaTeX{} \textit{to ScalaTion.} As seen earlier, many mathematical and statistical formulas can be expressed in ScalaTion. To this end, it would be convenient for users of the ScalaTion DSL if we implement a way to convert formulas written in \LaTeX{} to code that compiles with ScalaTion. This convenience extends beyond simply allowing users of the DSL to first write their formulas with \LaTeX. It also enables users to write their formulas in languages and environments (e.g., Maple) that support exportation to \LaTeX.

\item \textit{ScalaTion to} \LaTeX. Many times, it would be convenient if a user of the ScalaTion DSL could easily convert code written in ScalaTion to \LaTeX. (For instance, when preparing formulas for a paper.) As the syntax for both ScalaTion and \LaTeX is linear, it should be possible to easily parse a formula written in one and convert it to the other.

\item \textit{Prefix Operators via Compiler Plugins.} Unless Scala changes how it handles operator precedence and associativity, we need to work on ways to define such things for our Unicode-identified operators. This can be accomplished through the development of plugins for the Scala compiler. Possible implementations could include something as simple as regular expression substitution as a pre-processor phase or something as non-trivial as extending Scala's own lexical parser combinators.  We will also explore the language virtualization benefits of Rompf's \cite{rompf:lms} on Lightweight Modular Staging (LMS) in order to make these improvements easier to implement. Such work will help make the language both more familiar and easier to use by domain engineers. 

\end{itemize}

\acks

We would like to acknowledge the other ScalaTion group members for their contributions to the project: Jun Han, Maria Hybinette and Robert Davis.


\bibliographystyle{abbrv}


\bibliography{scalation_unicode}

\end{document}